\documentclass[prl,amsmath,aps]{revtex4-2} % twocolumn reprint, aps,,amssymb
\usepackage{graphicx}
\usepackage{epsfig}
\usepackage{color}
\usepackage{dcolumn}
\usepackage{bm}
\usepackage{float}
\usepackage{amsmath}
\DeclareMathAlphabet{\mathbbmsl}{U}{bbm}{m}{sl}
\makeatletter
\newsavebox{\@brx}
\newcommand{\llangle}[1][]{\savebox{\@brx}{\(\m@th{#1\langle}\)}%
	\mathopen{\copy\@brx\kern-0.5\wd\@brx\usebox{\@brx}}}
\newcommand{\rrangle}[1][]{\savebox{\@brx}{\(\m@th{#1\rangle}\)}%
	\mathclose{\copy\@brx\kern-0.5\wd\@brx\usebox{\@brx}}}
\makeatother
\begin{document}
\draft

\title{Quantized Hall current in topological nodal-line semimetal}

\author{Po-Hsin Shih$^{1}$, Thi-Nga Do$^{2}$, Godfrey Gumbs$^{1,3}$, Danhong Huang$^{4}\footnote{Corresponding author: {\em E-mail}: danhong.huang@us.af.mil }$, Hsin Lin$^{5}\footnote{Corresponding author: {\em E-mail}: nilnish@gmail.com }$, Tay-Rong Chang$^{2,6,7}\footnote{Corresponding author: {\em E-mail}: u32trc00@phys.ncku.edu.tw}$
}
\affiliation{$^{1}$ Department of Physics and Astronomy, Hunter College of the City University of New York,
695 Park Avenue, New York, New York 10065, USA \\
$^{2}$ Department of Physics, National Cheng Kung University, Tainan 701, Taiwan  \\
$^{3}$ Donostia International Physics Center (DIPC), P de Manuel Lardizabal, 4, 20018 San Sebastian, Basque Country, Spain \\
$^{4}$US Air Force Research Laboratory, Space Vehicles Directorate (AFRL$/$RVSU),\\
Kirtland Air Force Base, New Mexico 87117, USA \\
$^{5}$ Institute of Physics, Academia Sinica, Taipei 11529, Taiwan \\
$^{6}$ Center for Quantum Frontiers of Research and Technology (QFort),
Tainan 701, Taiwan \\
$^{7}$ Physics Division, National Center for Theoretical Sciences, Taipei
10617, Taiwan
}

\date{\today}

\begin{abstract}
Photocurrent acts as one of measurable responses of material to light, which has proved itself to be crucial for sensing and energy harvesting. Topological semimetals with gapless energy dispersion and abundant topological surface and bulk states exhibit exotic photocurrent responses, such as novel quantized circular photogalvanic effect observed in Weyl semimetals. Here we find that for a topological nodal-line semimetal (NLSM) with nodal ring bulk states and drumhead surface states (DSS), a significant photocurrent can be produced by an electromagnetic (EM) wave by means of the quantum Hall effect. The Hall current is enabled by electron transfer between Landau levels (LLs) and triggered by both the electric field and magnetic field components of an EM wave. This Hall current is physically connected to an unusually large quantum-Hall conductivity of the zeroth LLs resulting from quantized DSS. These LLs are found to be highly degenerate due to the unique band-folding effect associated with magnetic-field-induced expansion of a unit cell. Furthermore, we observe that the Hall current induced solely by an in-plane linearly-polarized EM wave becomes a quantized entity which allows for possible direct measurement of the DSS density in a topological NLSM.
This work paves a way toward designing high-magnetic-field-sensitivity detection devices for industrial and space applications, such as the development of self-detection of current-surge-induced overheating in electronic devices and accurate Earth's magnetic-anomaly maps for guiding a self-navigating drone or an aircraft.
\end{abstract}
\pacs{PACS:}
\maketitle

The response of solids to external fields, such as photocurrent, has been a central topic in solid state physics.
Photocurrent can be triggered by the photon absorption through various photoelectric effects like photovoltaic \cite{voltaic} and photogalvanic \cite{currentWeyl}. Optical quantum Hall effect (OQHE) has recently emerged as an alternative mechanism for the photocurrent generation. The optical quantum Hall current is generated by the charge pumping in the magnetic-field-induced Landau levels (LLs) in the systems with finite Hall conductance \cite{OQHE1, OQHE2, OQHE3}.
Up to now, the quantum Hall plateaus in the Terahertz regime have been investigated in a two-dimensional electron gas system \cite{OQHE1} and graphene \cite{OQHE2, OQHE3}. However, these OQHE require an external magnetic field, in addition to an incident light.
Though realization of OQHE engendered entirely by electromagnetic (EM) wave can be enabled by the recent advanced development of the modern materials fabrication and Terahertz light source technology, probing the OQHE still remains challenging because the induced current response is relatively weak and sensitive to the EM oscillation. Seeking materials with efficient EM wave-current conversion is highly desirable for sensing and energy harvesting.
\medskip

Topological semimetals (TSM) are promising materials to engender exotic photocurrent due to their special topological surface and bulk states. TSM are characterized by band crossing in the Brillouin zone (BZ) at or in the vicinity of the Fermi level ($E_F$). The conventional TSM can be classified into three different categories, namely, Dirac semimetal (DSM), Weyl semimetal (WSM), and nodal-line semimetal (NLSM) \cite{JPCM2016Weng}. The crossings of the bulk conduction and valence subbands of these systems form, respectively, the Dirac points \cite{JPCM2016Weng, DWSM2018Yang, DWSM2018Arm}, Weyl points \cite{JPCM2016Weng, DWSM2018Yang, DWSM2018Arm}, and one-dimensional (1D) nodal lines \cite{JPCM2016Weng, DWSM2018Yang, NLSM2019Chang, NLSM2020Song} in the BZ. The TSM support unique surface states, which have been identified for the Fermi-arc surface states in DSM \cite{fermiarc2015} and WSM \,\cite{TaAs2015}, and drumhead surface states (DSS) in NLSM \cite{DH2020Hosen, DH2020Muechler, DH2016Bian}. The photocurrent of TSM has been demonstrated unique, especially since the discovery of the quantized circular photogalvanic effect in WSM \cite{currentWeyl}, for which the photocurrent depends only on fundamental constants and the monopole charge of a Weyl node.
\medskip

Magnetic quantization is an important phenomenon which could be exploited for achieving the essential understanding of the topological behaviors in materials. This feature in TSM has enabled the realization of the Dirac fermions \cite{NP2014Fu, PRL2011Apa, PRL2010Cheng} and the chiral anomaly in DSM and WSM \cite{NC2018Yuan, Sci2015Xiong, NM2016Jia}. Although the magnetic quantization of NLSM was previously predicted \cite{oscillation2018Yang, Rhim2015} and discovered \cite{oscillation2018Li}, it has been limited to bulk LLs. Meanwhile the DSS LLs, which could bear crucial topological fingerprints of the system, remain largely unknown.
magnetic quantization is the part of QHE that describes the high-order response to external magnetic field of a system.
In general, a full theory of the electron response to higher order in external fields can be established via the Berry curvature and a first order correction to the band energy due to the orbital magnetic moment.
The derivation of the second-order field correction to the Berry curvature of Bloch electrons in external EM fields \,\cite{Gao2014} is essential for the study of response functions. So far, enormous attention has been devoted for the nonlinear effects of high-order electric field, such as the nonlinear Hall effect in the absence of magnetic field up to second order \cite{Sode2015} and third order \cite{Lai2021}.
Now, it is natural to ask if the high-order responses to both magnetic and electric fields can be addressed? and will it lead to any novel physical phenomena? Here, we answer these questions by presenting the optical quantum Hall current of high-order magnetic and electric fields in NLSM.
\medskip

The main achievements of this paper are threefold: (1) We observe the unique quantization phenomena of the NLSM DSS with unusually large, field-dependent surface QHC that have never been observed in the explored materials. By introducing a concept dealing with field-induced band folding, we uncover a fundamental understanding of the magnetic quantization mechanism in solids. We find that the degeneracies of LLs directly reflect the distribution of DSS within the folded BZ. (2) We derive semiclassically a general formula for the high-order current density and discover its relationship with the Berry curvature. This formula serves as an important basis for the investigation of photocurrent of materials. (3) We find a novel quantized signature in physics, which is the optical Hall current induced by an applied linearly polarized EM field on the DSS. The current response can be determined by only the fundamental constants and area of DSS.

\begin{figure}[h]
\centering
{\includegraphics[width=0.5\linewidth]{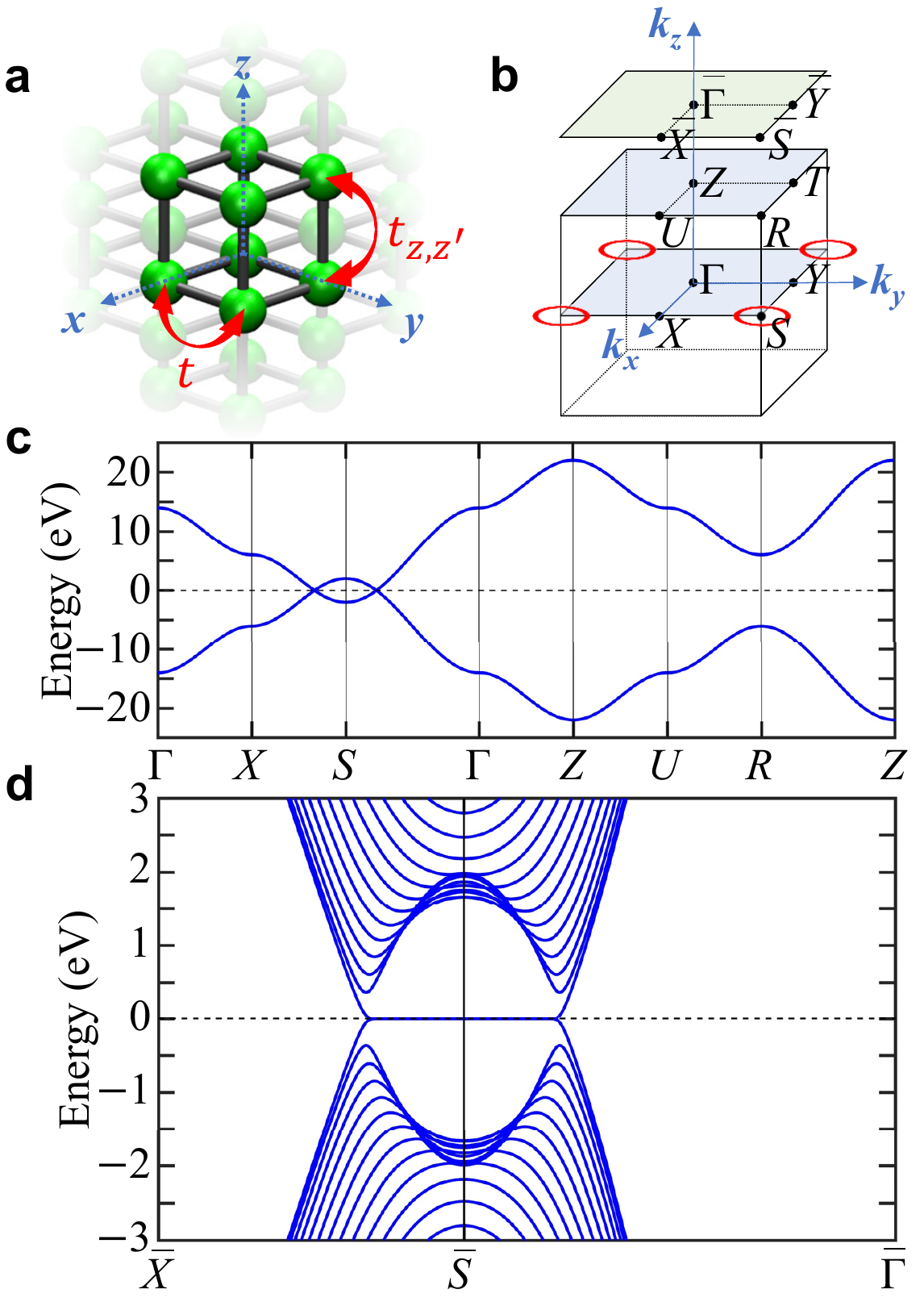}}
\caption{ {\bf a,} Crystal structure of a NLSM. {\bf b,} First Brillouin zone with high symmetry points for the bulk (lower) and slab (upper). For a 3D system, the first BZ is a cubic centered at $\Gamma$ point, while it changes to a square centered at $\bar{\Gamma}$ for a 2D slab. The red circles indicate the Dirac nodal rings of energy bands. {\bf c,} Bulk band structure along the high symmetry points. {\bf d,} Band structure of a 25-layer slab consists of (001) surface bands at zero energy and nearby bulk bands. }
\label{Fig1}
\end{figure}

We model the bulk NLSM by single atoms with two orbitals in the simple cubic unit cell, as seen in Fig.\,\ref{Fig1}a. The surface states can be formed in a slab containing multi-layers along the [001] axis. Figure\ \ref{Fig1}b shows the bulk BZ with high symmetry points on the $k_z$ = 0 plane [$\Gamma$, X, S, Y] and $k_z$ = $\pi$ plane [Z, U, R, T] as well as the (001)-projected surface BZ. We assume that the system has spin degeneracy. The minimum tight-binding Hamiltonian $2 \times 2$ matrix within the ($s, p_z$) basis for the bulk system can be written as

\begin{eqnarray}
H  =
\begin{bmatrix}
   t(f_1 + f_2) + t_z f^+_3 + \epsilon_0&
   t_{z'} f^-_3 \\
   -t_{z'} f^-_3&
    t(f_1 + f_2) + t_z f^+_3 - \epsilon_0
\end{bmatrix}
\label{eqn:1}
\end{eqnarray}
In Eq.\,\eqref{eqn:1}, $f_1 = e^{i\mbox{\boldmath$k$}.\mbox{\boldmath$r_1$}} + e^{-i\mbox{\boldmath$k$}.\mbox{\boldmath$r_1$}}$, $f_2 = e^{i\mbox{\boldmath$k$}.\mbox{\boldmath$r_2$}} + e^{-i\mbox{\boldmath$k$}.\mbox{\boldmath$r_2$}}$, and $f^{\pm}_3 = e^{i\mbox{\boldmath$k$}.\mbox{\boldmath$r_3$}} \pm e^{-i\mbox{\boldmath$k$}.\mbox{\boldmath$r_3$}}$ are the phase terms with $\mbox{\boldmath$k$}$ the wave vector and $\mbox{\boldmath$r_{1,2,3}$}$ the unit vectors along $x$, $y$, $z$ directions. The hopping integral $t$ = $-2$ eV is for the horizontal interactions, while $t_{z}$ = 2 eV and $t_{z'}$ = 1 eV are for the interactions between the same and different orbital domains in the vertical direction, respectively. $\epsilon_0 = 10$ eV is the site energy.
\medskip

The calculated bulk band structure along the high symmetry points is shown in Fig.\,\ref{Fig1}c. The intersection of the conduction and valence bands forms a nodal ring encircling S on the $k_z$ = 0 plane, giving the band crossing along the S-X and S-$\Gamma$ directions in Fig.\,\ref{Fig1}c. On the contrary, the conduction and valence bands disperse apart elsewhere in the whole first BZ. The Dirac lines feature four-fold degeneracies, associated with the band crossing and equivalence between spin states. The slab band structure, shown in Fig.\,\ref{Fig1}d, consists of DSS around the $\bar{S}$ point and numerous highly dispersive bulk bands. The DSS are bounded by the projected nodal ring and almost dispersionless energy band. They are four-fold degenerate where both the electron and hole surface states coexist with spin degeneracy.

\begin{figure}[h]
\centering
{\includegraphics[width=0.5\linewidth]{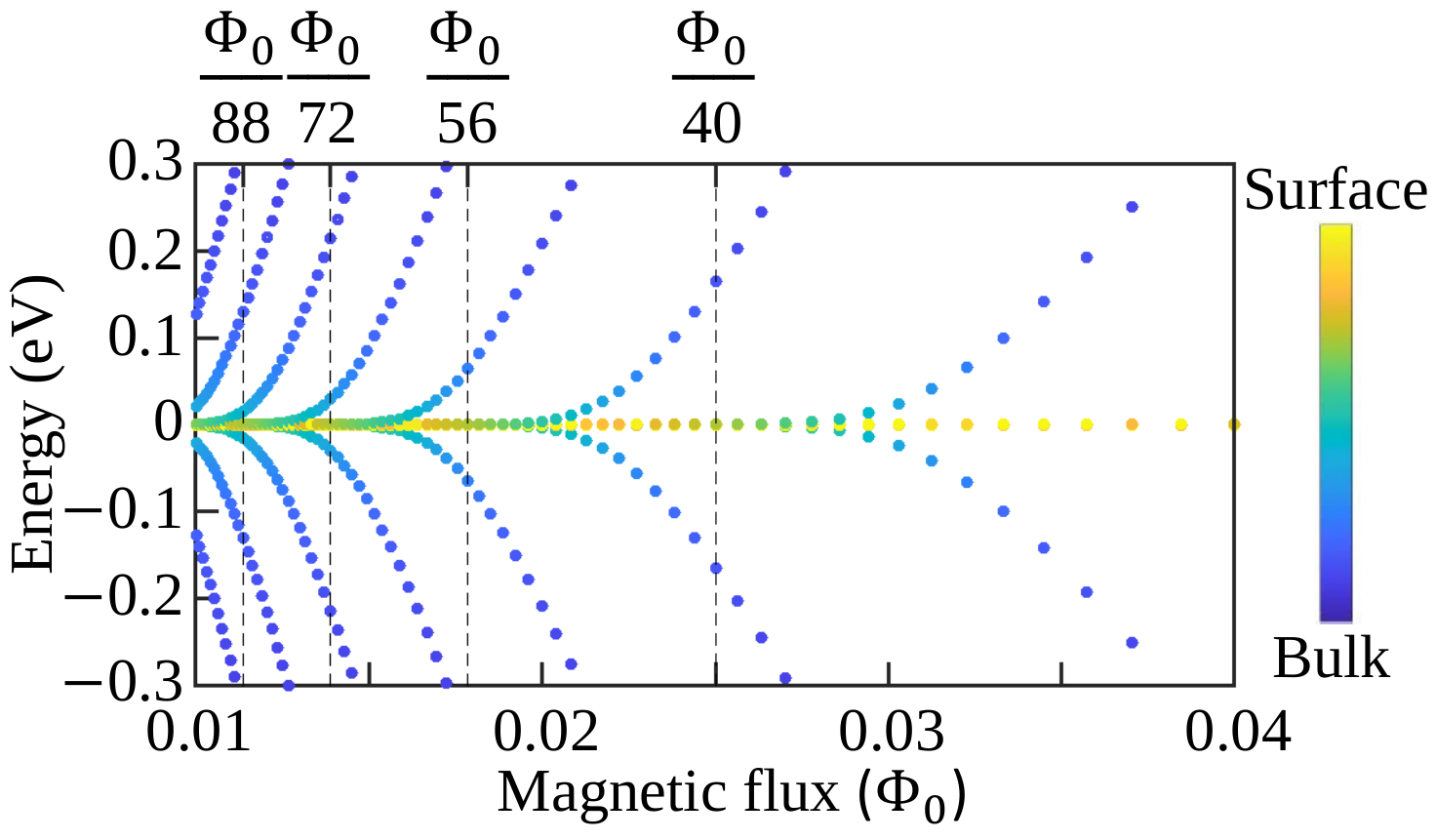}}
\caption{The magnetic-flux-dependent LL energy spectrum for a $25$-layer slab. The surface and bulk states are illustrated by yellow and blue dots, respectively. The surface spectral weights are defined as the states dominated by outer three layers of slab.}
\label{Fig2}
\end{figure}

\medskip

When a 2D condensed-matter system is subjected to a perpendicular magnetic field $\mbox{\boldmath$B$}$ = (0,0,B), electrons follow quasi-classical cyclotron motion, thus electronic states are quantized into LLs. The application of magnetic field changes the lattice periodicity so that the primitive unit cell is extended along the $x$ direction. Particularly, the field-induced Peierls phase of the form $G_{R} = (2\pi/\Phi_{0})\int\limits_R^r \mbox{\boldmath$A$}\cdot d\mbox{\boldmath$\ell$}$ needs to be included in the Hamiltonian \cite{quantization}. Here, $\mbox{\boldmath$A$}=(0,Bx,0)$ is the vector potential in the Landau gauge and $\Phi_{0}=h/e$ is the flux quantum. The Peierls phases are repeated periodically along with the extended unit cell in the lattice when the total magnetic flux equals to $\Phi_0$.
\medskip

The $\Phi$-dependent spectrum of DSS LLs is demonstrated in Fig.\,\ref{Fig2}. $\Phi = B {\cal S}$ is the magnetic flux per unit cell with ${\cal S}$ being the area of the primitive unit cell in real space. We observe that the quantized LLs of DSS behaves similarly to the zeroth LLs of graphene and surface states of topological insulator (TI). For these Dirac systems \cite{graphene2005Nov, graphene2005Gus, NP2014Fu}, it was previously shown that the flat zeroth LLs at the Dirac-point energy are independent of magnetic field strength. The LLs at higher energy, which arise from the linear band, acquire the square root dependence on both the LL index and the magnetic field. For NLSM, the magnetic quantization of DSS yields a group of field-independent and non-dispersive zeroth LLs, which only exist at $E_F$ = 0. Note that, the surface LLs of NLSM are mainly produced from DSS, in contrast to the zeroth LLs of graphene and surface states of TI which are quantized partially from nearby states. With the increase of magnetic field, the zeroth LLs gradually deform into bulk LLs at critical fields $\Phi_{N_L} = \frac{A_{DSS}}{A_{0,BZ}} \frac{\Phi_0}{N_L - 1/2}$, in which $N_L$ is the number of degeneracies of the zeroth LLs and $A_{DSS}/A_{0,BZ}$ defines the ratio of the DSS area over the BZ area at zero $\mbox{\boldmath$B$}$ field. Consequently, the DSS LL degeneracies decrease by the number of peeled-off zeroth LLs. In fact, the surface LLs become highly degenerate only at low fields. This characteristic is unique for the DSS which, to our knowledge, has never been observed in any other explored materials, and it plays a critical role in transport properties. The LLs merging mechanism can be understood well through the band-folding effect \cite{Sym2017Topp, bandfolding}, which is accompanied by an enlarged real-space unit cell and a reduced size of the first BZ (details in Section I of Supplementary materials).
\medskip

\begin{figure}[h]
\centering
{\includegraphics[width=0.5\linewidth]{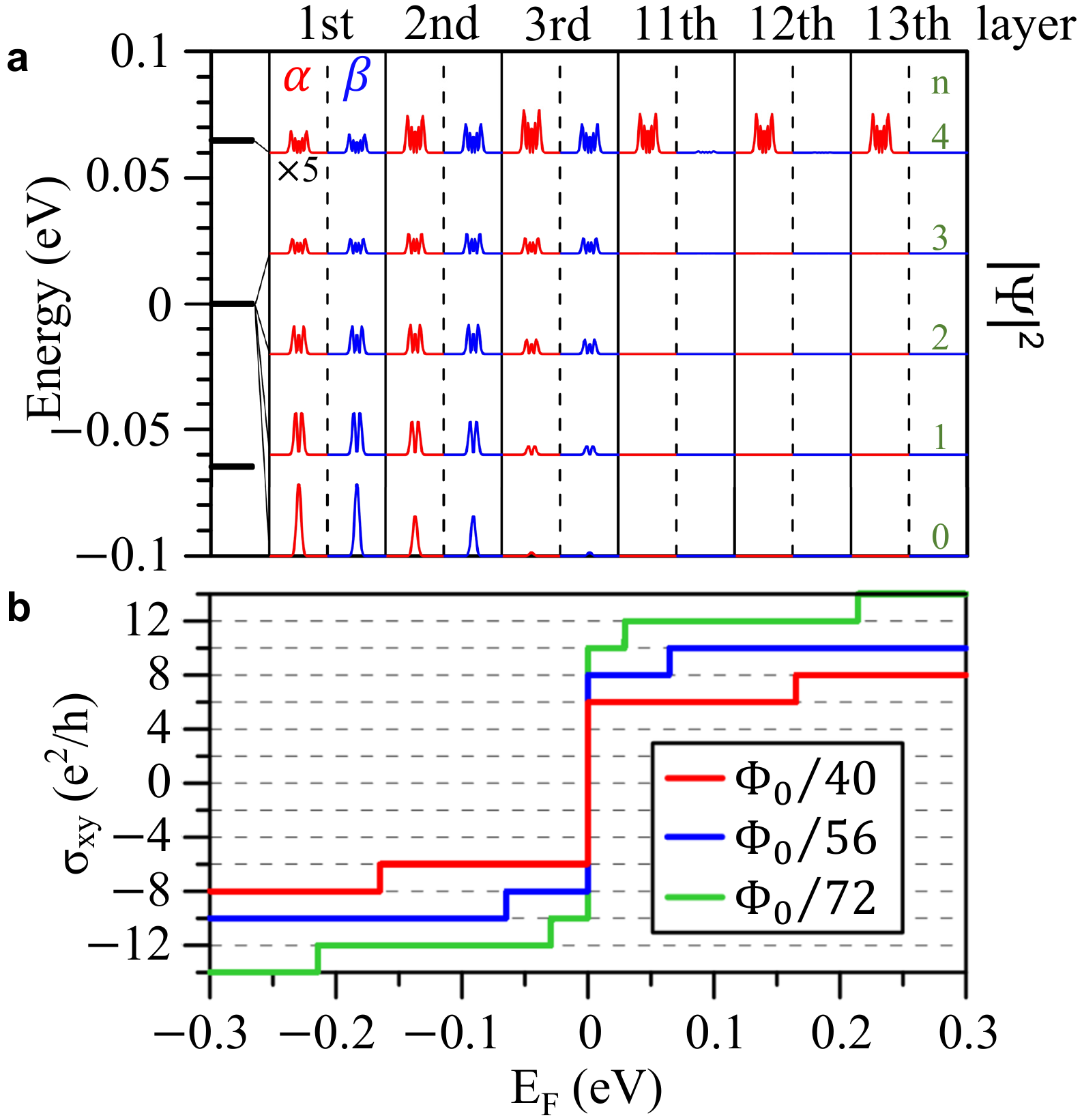}}
\caption{{\bf a,} Selected spatial dependence of probability functions for LLs at low energies in a $25$-layer slab. Here, $\alpha$ and $\beta$ label two different orbitals. {\bf b,} $E_F$ dependence of QHC at various $\Phi$'s.}
\label{Fig3}
\end{figure}

The probability function $|\Psi|^2$, defined as the square of the magnitude of the wave function, is useful for identifying LLs. Figure\,\ref{Fig3}a depicts  $|\Psi|^2$ for both surface and bulk LLs for $\Phi$ = $\Phi_0$/56 on both $\alpha$ and $\beta$ orbitals. Here, $|\Psi|^2$ exhibits well-behaved oscillatory modes, and the number of zero nodes determines the corresponding LL index $n$. Since each pair of conduction and valence LLs acquires the same $|\Psi|^2$, there are only four different oscillation modes, labeled by $n = 0,\,1,\,2,\,3$ for the eight-fold-degenerate zeroth LLs. Based on $|\Psi|^2$ of the LLs at zero energy, it is clear that these modes are dominated by the outer-three layers of the slab, i.e., they are the surface states. At higher energies, LLs (e.g., $n = 4$ LL) have nonvanishing $|\Psi|^2$ on all layers, implying their bulk properties. Therefore, the distribution property of $|\Psi|^2$ can be utilized to select out the surface LLs from bulk LLs in the system. This is considered as a key step in studying QHC of the DSS.

\medskip
The QHE, one of the most essential electronic transport signatures for topological materials, exists a robust connection with magnetic quantization. The QHC is well quantized when $E_F$ lies in the gap between two LLs. The $E_F$-dependent QHC, shown in Fig.\,\ref{Fig3}b, is calculated by employing the Kubo formula in the form

\begin{eqnarray}
\sigma_{xy} &=& \frac {ie^2 \hbar} {S}
\sum\limits_{n} \sum\limits_{n^{\prime} \neq n} (f_{n} - f_{n^{\prime}})
\frac {\langle \Psi_n  |\dot{\mbox{\boldmath$u$}}_{x}| \Psi_{n^{\prime}}\rangle  \langle \Psi_{n^{\prime}} |\dot{\mbox{\boldmath$u$}}_{y}|\Psi_n \rangle} {(E_{n}-E_{n^{\prime}})^2 + \Gamma_0 ^2} \,.
\label{eqn:2}
\end{eqnarray}
In this notation, $E_{n}$ is the LL energy and $|\Psi_{n}\rangle$ is the corresponding $n$th-LL wave function. They are evaluated from the tight-binding Hamiltonian in Eq.\,\eqref{eqn:1} and illustrated in Fig.\,\ref{Fig3}a. $\dot{\mbox{\boldmath$u$}}_{x,y}$ are the velocity operators, $f_{n}$ is the Fermi-Dirac distribution function, and $\Gamma_0$ ($\sim$ 1 meV) is the broadening factor. The calculated QHC displays the step features in which the plateaus correspond to vertical transition from occupied to unoccupied LLs. We found an unusually large QHC step for the zeroth LLs with high degeneracy via the relationship $\sigma_{xy} = C$($e^2/h$)=2$N_L$($e^2/h$), which implies the huge Chern number $C$ as well as enormous Berry curvature. On the contrary, the steps of 2$e^2/h$ are obtained for the bulk LLs due to the two-fold spin degeneracy. Such substantial variation of the QHC at a certain energy range has never been observed in the explored materials. For NLSM, the unique distribution of the DSS leads to the relation
\begin{eqnarray}
N_L \cong \frac{A_{DSS}}{A_{B,BZ}} = \frac{A_{DSS}}{A_{0,BZ}} \frac{\Phi_0}{B\cal S},
\label{eqn:3}
\end{eqnarray}
in which, $A_{B,BZ}$ is the folded BZ area under $\mbox{\boldmath$B$}$ field. This approximation is made within the limit of weak $\mbox{\boldmath$B$}$ field (details in Section I of Supplementary Material).
In fact, the occupation of DSS in the first BZ can be manipulated by tuning the tight-binding parameters. Explicitly, by increasing the vertical hopping terms $t_z$ and $t_{z'}$, the area of DSS is enhanced accordingly, leading to the change of critical fields for LLs merging and QHC steps. Our modeling and computations reveal similar features in energy bands, LL spectra, and QHC at $E_F$ = 0 for various sets of chosen parameters.

\begin{figure}[h]
\centering
{\includegraphics[width=0.5\linewidth]{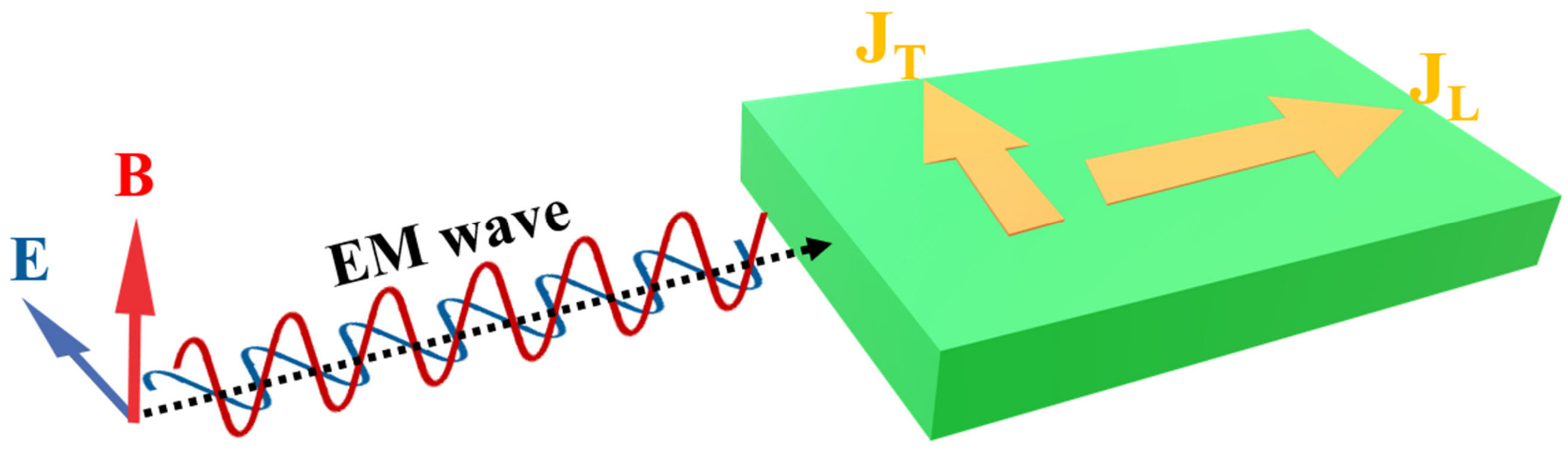}}
\caption{Visual illustration of longitudinal ($\mbox{\boldmath$J_L$}$) and transverse ($\mbox{\boldmath$J_T$}$) currents in a NLSM under an EM field.
}
\label{Fig4}
\end{figure}

\medskip
When a 2D system is subject to a linearly polarized EM field, there will occur a current across the material as a result of the OQHE or photoelectric effects \cite{OQHE1, OQHE2, OQHE3, voltaic, currentWeyl}. The high-order anomalous equilibrium (Berry-Hall) current component can be expressed semiclassically as (see Section II of Supplementary Materials)

\begin{eqnarray}
\nonumber
\mbox{\boldmath$j$}_1(t\,\vert\,E,B)&=&\frac{1}{\hbar}\int\frac{d^2\mbox{\boldmath$k$}}{(2\pi)^2}{f_0[\varepsilon_n^{(0)}(\mbox{\boldmath$k$}\,\vert\,B)-\mu_e]}
\\
\nonumber
&\times& \{\mbox{\boldmath$\nabla$}_{\bf k}[\varepsilon^{(0)}_n(\mbox{\boldmath$k$}\,\vert\,B)+e\mbox{\boldmath$E$}(t)\cdot\mbox{\boldmath${\cal A}$}^{(0)}_n(\mbox{\boldmath$k$}\,\vert\,B) \\
\nonumber
&+e&\mbox{\boldmath$E$}(t)\cdot [\overleftrightarrow{\mbox{\boldmath${\cal G}$}_n}(\mbox{\boldmath$k$}\,\vert\,B)\cdot \mbox{\boldmath$E$}(t)]]
+e\mbox{\boldmath$E$}(t)\times\mbox{\boldmath$\Omega$}^{(0)}_n(\mbox{\boldmath$k$}\,\vert\,B) \\
&+e&\mbox{\boldmath$E$}(t) \times [\mbox{\boldmath$\nabla$}_{\bf k} \times \overleftrightarrow{\mbox{\boldmath${\cal G}$}_n}(\mbox{\boldmath$k$}\,\vert\,B)\cdot \mbox{\boldmath$E$}(t)]\}.
\label{eqn:4}
\end{eqnarray}
Here, $\varepsilon_n^{(0)}(\mbox{\boldmath$k$}\,\vert\,B)$ represents the LL energies, $\overleftrightarrow{\mbox{\boldmath${\cal G}$}_n}(\mbox{\boldmath$k$}\,\vert\,B)$ is the Berry-connection polarizability tensor, $\mbox{\boldmath${\cal A}$}^{(0)}_n(\mbox{\boldmath$k$}\,\vert\,B)$ is the unperturbed Berry connection and $\mbox{\boldmath$\Omega$}^{(0)}_n(\mbox{\boldmath$k$}\,\vert\,B)$ is the unperturbed Berry curvature of Bloch electrons. The anomalous thermal-equilibrium current includes both the parallel ($\mbox{\boldmath$J_L$}$) and perpendicular ($\mbox{\boldmath$J_T$}$) components respect to the direction of the EM field, referring to Fig. 4. They are associated with the quantized transverse conductivity and continuous longitudinal conductivity, for unique all-electron thermal-equilibrium transports.
Note that both $\varepsilon_n^{(0)}(\mbox{\boldmath$k$}\,\vert\,B)$ and $\overleftrightarrow{\mbox{\boldmath${\cal G}$}_n}(\mbox{\boldmath$k$}\,\vert\,B)$ are independent of $k$, thus the first and third and fifth terms are vanishing (see Supplementary Material). The second term can not be physically observable since the Berry connection $\mbox{\boldmath${\cal A}$}^{(0)}_n(\mbox{\boldmath$k$}\,\vert\,B)$ is a gauge-dependent variable.
In general, the substantial $B$-dependent Berry curvature $\mbox{\boldmath$\Omega$}^{(0)}_n(\mbox{\boldmath$k$}\,\vert\,B)$ plays the key role in determining the current response of the system. On the other hand, $\mbox{\boldmath$\Omega$}^{(0)}_n(\mbox{\boldmath$k$}\,\vert\,B)$ only has contribution to the longitudinal current. As a matter of fact, the $B$-dependent anomalous optical Hall current flowing in NLSM is dominated by $\mbox{\boldmath$J_L$}$, which can be written as (details in Section III of Supplementary Materials)
\begin{eqnarray}
\mbox{\boldmath$J_L$} &=& \sigma_{xy}E\vec{n}
\cong 2\frac{A_{DSS}}{A_{0,BZ}} \frac{\Phi_0}{\cal S} \frac{c e^2}{h}\vec{n}.
\label{eqn:5}
\end{eqnarray}
In this notation, $\vec{n}$ is the direction of the incident EM field. It is clear that $\mbox{\boldmath$J_L$}$ depends only on the fundamental constants ($e$, $h$, $c$, $\Phi_0$) and the intrinsic characteristics of the NLSM sample.
In other words, the EM wave-generated Hall current in NLSM is a quantized signature.
So far, the quantized current response has also been predicted in WSM, in particular, the injection current depends only on the fundamental constants and the topological charge of Weyl nodes \cite{currentWeyl}. Such current is induced by the circular photogalvanic effect under a circularly polarized light. The condition for quantization of current response in WSM is the breaking of inversion and mirror symmetries, different from the time-reversal symmetry breaking in NLSM under QHE.
\medskip

Experimentally, a current density of about $\frac{A_{DSS}}{A_{0,BZ}} \frac{ec}{\cal S}$ can be observed in NLSM when a suitable sample is placed under an EM field. This is up to five orders larger than that of graphene under the same condition \cite{currentgraphene}.
Such topologically protected quantized Hall current will remain unchanged regardless of the incident in-plane linearly polarized EM field, as long as the DSS is quantized into LLs.
For the experiment setup, the sample needs to be sufficiently large for the $\mbox{\boldmath$B$}$-induced cyclotron motion of electrons to be formed within the magnetic length $l_B = \sqrt{\hbar/eB}$.
For example, an EM field with $E$ = 3 $\times$ 10$^7$ (V/m) and $B$ = 0.1 T requires a minimum lattice sample of 400 $\times$ 400 nm$^2$ for the observation of current \,\cite{expj}.
From an application perspective, the significant and quantized optical Hall current can be put to technological applications and designing high-sensitivity detection devices. The robust connection between the current density and DSS enables the direct measurement of the density of DSS in topological NLSM.
Furthermore, the strong dependence of the surface QHC on the $\mbox{\boldmath$B$}$ field paves a way toward the development of $\mbox{\boldmath$B$}$-sensitive detectors for industrial and space applications.
In particular, such a concise physics picture can be employed for developing anomalous-Hall-effect based compact and ultra-sensitive magnetometers \cite{app} in measuring a weak magnetic field. Consequently, the anomaly measurement of Earth's magnetic field, aided by afterward machine-learning processing for the enhancement of edge contrast, enables extensively extracting ground-surface profile data quickly and accurately. This processes can be followed further by matching obtained ground-surface profile to available accurate magnetic-anomaly maps for guiding a self-navigating drone or an aircraft with reasonable accuracy in areas without accessible GPS signals as well as applicable to self-detection of overheating due to current surge in automobiles or power-supply networks based on Faraday effect.
\medskip

In conclusion, we have shown that the novel EM wave-induced QHE of the topological NLSM is a quantized response based on the novel magnetic quantization of DSS and its connection with field-induced band folding. We found the unusually large, field-dependent surface QHC induced by the extremely high LL degeneracy. We set an important groundwork for the study of photocurrent by deriving a general semiclassical formula for the high-order photocurrent. This work has established a new hallmark for studies of NLSM, which could play a critical role in next-generation technology and high-performance device applications.

\section*{Acknowledgement(s)}
D.H. would like to acknowledge the financial support from the Air Force Office of Scientific Research (AFOSR). G.G. would like to acknowledge the support from the Air Force Research Laboratory (AFRL) through Grant No. FA9453-21-1-0046. T.-R.C. was supported by the Young Scholar Fellowship Program from the MOST in Taiwan, under a MOST grant for the Columbus Program, No. MOST110-2636-M-006-016, NCKU, Taiwan, and the National Center for Theoretical Sciences, Taiwan. Work at NCKU was supported by the MOST, Taiwan, under Grant No. MOST107-2627-E-006-001 and the Higher
Education Sprout Project, Ministry of Education to the Headquarters of University Advancement at NCKU. T.-N.D. would like to thank the MOST of Taiwan for the support through Grant No. MOST111-2811-M-006-009.

\end{document}